\documentclass[letterpaper, 10 pt, conference]{ieeeconf}  %
\IEEEoverridecommandlockouts                              

\usepackage{array}
\usepackage{textcomp}
\usepackage{stfloats}
\usepackage{url}
\usepackage{verbatim}
\usepackage{times}

\usepackage{bm}
\usepackage[numbers]{natbib}
\usepackage{multicol}
\usepackage{xcolor}
\usepackage{paralist}
\usepackage{comment}
\usepackage{footnote}
\usepackage{graphicx}
\usepackage{multirow}
\usepackage{multicol}
\usepackage{slashbox}
\usepackage[most]{tcolorbox}
\usepackage{amsmath, amsfonts, amssymb}
\usepackage[pagebackref=false,breaklinks=true,colorlinks=true, citecolor=magenta,linkcolor=orange,urlcolor=blue]{hyperref}

\newtheorem{assumption}{Assumption}

\newtheorem{theorem}{Theorem}

\def\BibTeX{{\rm B\kern-.05em{\sc i\kern-.025em b}\kern-.08em
		T\kern-.1667em\lower.7ex\hbox{E}\kern-.125emX}}
\usepackage{balance}

\begin{document}
	\title{\LARGE \bf 
	Fast Whole-Body Strain Regulation in Continuum Robots 
	} 
	
	\author{Lekan Molu 
		\\
		\href{https://github.com/robotsorcerer/DCM/tree/SPT}{\textcolor{blue}{\texttt{https://github.com/robotsorcerer/DCM/tree/SPT}}}
	\thanks{The author is with Microsoft Research NYC {\tt \small lekanmolu@microsoft.com}. 
}}
	
	\maketitle


\newcommand{\lb}[1]{\textcolor{light-blue}{#1}}
\newcommand{\bl}[1]{\textcolor{blue}{#1}}

\newcommand{\maybe}[1]{\textcolor{gray}{\textbf{MAYBE: }{#1}}}
\newcommand{\inspect}[1]{\textcolor{blue}{\textbf{CHECK THIS: }{#1}}}
\newcommand{\more}[1]{\textcolor{red}{\textbf{MORE: }{#1}}}
\renewcommand{\figureautorefname}{Fig.}
\renewcommand{\sectionautorefname}{$\S$}
\renewcommand{\equationautorefname}{equation}
\renewcommand{\subsectionautorefname}{$\S$}
\renewcommand{\chapterautorefname}{Chapter}

\newcommand{\cmt}[1]{{\footnotesize\textcolor{red}{#1}}}
\newcommand{\lekan}[2]{{\footnotesize\textcolor{purple}{LM: #1}}{#2}}
\newcommand{\gilbert}[2]{{\footnotesize\textcolor{blue}{GB: #1}}{#2}}
\newcommand{\audrey}[2]{{\footnotesize\textcolor{magenta}{AS: #1}}{#2}}
\newcommand{\ames}[2]{{\footnotesize\textcolor{pink}{AA: #1}}{#2}}
\newcommand{\todo}[1]{\textcolor{cyan}{TO-DO: #1}}
\newcommand{\review}[1]{\noindent\textcolor{red}{$\rightarrow$ #1}}
\newcommand{\response}[1]{\noindent{#1}}
\newcommand{\stopped}[1]{\color{red}STOPPED HERE #1\hrulefill}

\newcommand{\linkToPdf}[1]{\href{#1}{\blue{(pdf)}}}
\newcommand{\linkToPpt}[1]{\href{#1}{\blue{(ppt)}}}
\newcommand{\linkToCode}[1]{\href{#1}{\blue{(code)}}}
\newcommand{\linkToWeb}[1]{\href{#1}{\blue{(web)}}}
\newcommand{\linkToVideo}[1]{\href{#1}{\blue{(video)}}}
\newcommand{\linkToMedia}[1]{\href{#1}{\blue{(media)}}}
\newcommand{\award}[1]{\xspace} %

\newcounter{mnote}
\newcommand{\marginote}[1]{\addtocounter{mnote}{1}\marginpar{\themnote. \scriptsize #1}}
\setcounter{mnote}{0}
\newcommand{\ie}{i.e.\ }
\newcommand{\eg}{e.g.\ }
\newcommand{\cf}{cf.\ }
\newcommand{\yes}{\checkmark}
\newcommand{\no}{\ding{55}}

\newcommand{\flabel}[1]{\label{fig:#1}}
\newcommand{\seclabel}[1]{\label{sec:#1}}
\newcommand{\tlabel}[1]{\label{tab:#1}}
\newcommand{\elabel}[1]{\label{eq:#1}}
\newcommand{\alabel}[1]{\label{alg:#1}}
\newcommand{\fref}[1]{\cref{fig:#1}}
\newcommand{\sref}[1]{\cref{sec:#1}}
\newcommand{\tref}[1]{\cref{tab:#1}}
\newcommand{\eref}[1]{\cref{eq:#1}}
\newcommand{\aref}[1]{\cref{alg:#1}}

\newcommand{\bull}[1]{$\bullet$ #1}
\newcommand{\argmax}{\text{argmax}}
\newcommand{\argmin}{\text{argmin}}
\newcommand{\mc}[1]{\mathcal{#1}}
\newcommand{\bb}[1]{\mathbb{#1}}

\def\tidx{t}
\def\reline{\mathbb{R}}
\def\ren{\mathbb{R}^n}
\newcommand{\Note}[1]{}
\renewcommand{\Note}[1]{\hl{[#1]}}  


\def\kau{\mc{K}}
\def\particle{\bm{x}}
\def\zero{\bm{0}}
\def\hcal{\bm{\mc{H}}}
\def\materialresponse{\bm{G}}
\def\orthoggroup{{\textit{SO}}(3)}
\def\liegroup{{\mathbb{SE}}(3)}
\def\liealgebra{\mathfrak{se}(3)}
\def\identity{\bm{I}}
\def\adj{\text{ad}}
\def\Adj{\text{Ad}}
\def\pos{p}
\def\rot{R}
\def\gain{\bm{K}}
\def\error{\bm{e}}
\def\zee{\bm{z}}
\def\alp{\bm{\alpha}}
\def\titi{\bm{\theta}}
\def\alfa{bm{\alpha}}
\def\core{\text{fast}}
\def\pert{\text{slow}}
\def\vinput{\bm{\nu}}
\def\ufast{u_f}
\def\abscissa{\bm{X}}
\def\conf{\bm{g}}
\def\torque{\bm{u}}
\def\ones{\bm{1}}
\def\strain{\xi}
\def\twist{\eta}
\def\gencoord{\bm{q}}
\def\hilbert{{H}}
\def\jacob{\bm{J}}
\def\pde{p.d.e.}
\def\cref{cf.\ }
\def\perturb{\epsilon}
\def\massmat{\bm{M}}
\def\forcemat{\bm{F}}
\def\cormat{\bm{C}}
\def\dragmat{\bm{D}}
\def\Nmat{\bm{N}}
\def\masscom{\bm{\mc{M}}}
\def\masscore{\masscom^{\text{core}}}
\def\masspert{\masscom^{\text{pert}}}
\def\lyapweight{\varepsilon}

	\begin{abstract}
We propose reaching steps towards the real-time strain control of multiphysics, multiscale continuum soft robots. 
To study this problem fundamentally, we ground ourselves in a model-based control setting enabled by mathematically precise dynamics of a soft robot prototype. Poised to  integrate, rather than reject, inherent mechanical nonlinearity for embodied  compliance, we first separate the original robot dynamics into two separate subdynamics --- aided by a perturbing time-scale separation parameter. Second, we prescribe a set of stabilizing nonlinear backstepping controllers for regulating the resulting subsystems' strain dynamics. Third, we study the interconnected singularly perturbed system by analyzing and establishing its stability. Fourth, our theories are backed up by fast numerical results on a single arm of the Octopus robot arm. We demonstrate strain regulation to equilibrium, in a significantly reduced time, of the whole-body reduced-order dynamics of infinite degrees-of-freedom soft robots. This paper communicates our thinking within the backdrop of embodied intelligence: it informs our conceptualization, formulation, computational setup, and  yields improved control performance  for the nonlinear control of infinite degrees-of-freedom soft robots.
\end{abstract}

	\section{Introduction}
\label{sec:intro}

Soft manipulators, inspired by the functional role of  living organisms' soft tissues, provide better compliance and configurability compared to their rigid counterparts. In proof-of-concept studies and in certain real-world cases, they have found applications in delicate 6D dexterous bending and whole-arm manipulation tasks~\cite{Godage}, minimally invasive surgery in tight spaces~\cite{LekanIROS17, LekanICRA20}, inspection~\cite{Mehling}, and assistive rehabilitation~\cite{ConorSoroGlove, ConorLamb} tasks, where otherwise stiff and rigid robot configurations possess worse stiffness-to-weight ratios and manipulability. Despite their attractiveness, rigid robots are still the go-to mechanism in many automation tasks today. How can we bridge this divide for soft robot adoption in everyday automation? In this paper, we argue that a sustained research effort for developing real-time computational tools for interaction modeling and control will be the key to wide adoption.

\begin{figure}[tb!]
	\centering
		\includegraphics[width=0.85\columnwidth]{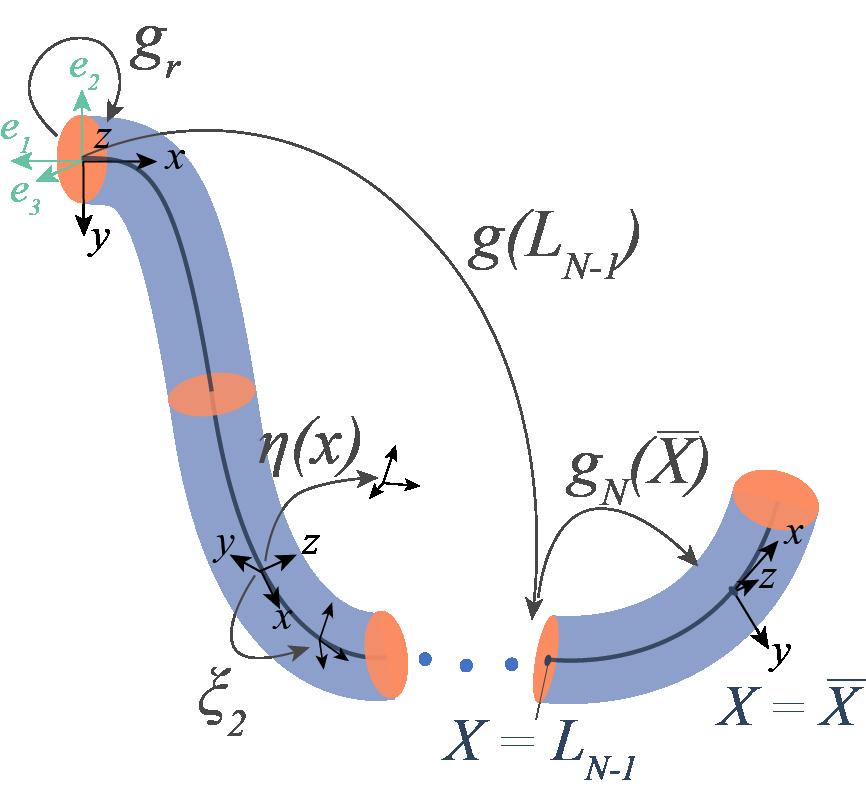} 
	%
	\caption{
		\footnotesize{Simplified configuration of an Octopus arm, reprinted from \citet{LekanSoRoPD}}. 
	}
	\label{fig:pcs_kine}. 
\end{figure}

Soft robots are multiphysics systems that generate physically heterogeneous interactions from muscle activation to contact and adhesion with the environment in an embodied intelligence fashion~\cite{Zambrano}. Embodied intelligence stipulates that rather than reject external mechanical processes that impede performance, a robot should leverage its shape, geometry of components, along with constraints in the external environment to achieve its desired configuration. Given that the nonlinear deformation of soft robots occur at multiple scales: from millimeters (in their continuum deformation characterization) to meters (in their overarching compliance strategy), we are poised with the fast and precise control of soft robots. To systematically dissect the problem, we focus on model-based control methods. This is attractive since the long time scales required to computationally resolve models and control has been a drawback for their ubiquitous adoption. 

We take a holistic approach that includes modeling, applied mathematics and control, and fast scientific computing schemes to solve the multiscale problem constrained by the robot's multiphysics. Being a continuum phenomenon, the default machinery for soft robot analyses are nonlinear partial differential equations (PDEs). However, nonlinear PDE theory is tedious and computationally intensive for realizing computationally fast and compliant behavior in soft robots. There are notable strides in   reduced-order, finite-dimensional mathematical models that induce tractability in continuum models. A non-exhaustive list range from morphoelastic filament theory~\cite{GorielyMorphoelastic, KuhlGorielyFilament, gazzola2018forward}, to generalized Cosserat rod theory~\cite{RubinCosserat, CosseratBrothers}, the constant curvature model~\cite{Godage}, the piecewise constant curvature model~\cite{PCC-Robert, RSSPCC}, and ordinary differential equations-based discrete Cosserat model~\cite{RendaICRA16, RendaTRO18}.  

To study the problem at hand, we leverage~\cite{RendaTRO18}'s kinetic model in  grounding the layered multirate control scheme~\cite{NikMatniLCA}  of the various interconnected physics components of a soft robot prototype. In this sentiment, we take the view of reduced order modeling and control with singular perturbation techniques~\cite{SPTBook}. Discretizing the continuum into piecewise constant strain sections~\cite{RendaTRO18}, we consider regions where the robot's activation influences its mass density the most as the fast subsystem to be controlled on a finer scale. The remaining microstructures on the robot are considered the slower subsystem which can be solved at a  much coarser resolution. This enables us to devise a tractable mathematical scheme for separating the system dynamics into two separate sub-dynamical systems, controllable at different time scales, to improve computational time and accurate strain regulation. To encourage resilience and improve runtime, we sidestep linear control methods~\cite{PSatID, LekanSoRoPD} and opt for nonlinear control whilst exploiting interprocess communication on a modern GPU and its host CPU. The motivation is for the robot to utilize, not discard, its inherent  mechanical nonlinear feedback in achieving control compliance whilst  improving computational time. 

\noindent \textbf{Contributions}: 
Our contributions are as follows:
\begin{itemize}
	\item we separate the robot dynamics into separate time scales by manipulating the governing dynamics equations with a perturbation parameter;
	\item we then devise separate nonlinear controllers for either subdynamics, each operating at different time resolutions on separate GPU and host CPU threads;
	\item between the two separated subdynamics,  an asynchronous communication scheme enables  passing dynamics and control computational data from one thread to the other -- the subdynamics and controller of the other system are ``frozen" within the other subsystem's control and dynamics thread -- we do not freeze the other process itself;
	%
	%
	\item a multi-rate sampling of state measurements asynchronously controls each subsystem: a fast sampling of the fast state variable is employed in a fast nonlinear backstepping controller and a slow-sampling of the slow state variable is employed in a slow backstepping controller. There is not a stringent requirement for communication between both subsystems so that the overall controller takes the form of a decentralized one;
	\item we achieve a faster computational time for control compared to previously reported results~\cite{HierarchicalGazzolla, LekanSoRoPD}.
\end{itemize}  
Our formulation avoids the empirical hierarchical computational schemes typically employed on soft robot bodies such as~\citet{HierarchicalGazzolla}. While in a way our contribution adheres to this bio-inspired hierarchical computational scheme,a layered modeling and control scheme from a rigorous dynamical systems viewpoint enables us to preserve stability guarantees to the computational scheme. This allows the negligence of
\begin{inparaenum}[(i)]
	\item  parasitic parameters which otherwise complicate system model;
	\item extraneous minute time constants, and mass densities etc; and 
	\item the overparameterization caused by sensitive neural network (and hence non-interpretability of) models used for the high-level controllers in bio-inspired models such as~\cite{HierarchicalGazzolla}.
\end{inparaenum}

The rest of this paper is structured as follows: background and theoretical machinery are described in \S \ref{sec:prelims}; \S \ref{sec:fast-slow} introduces the singularly perturbed dynamics framework and in \S \ref{sec:hcs}, we prescribe the layered dynamics and backstepping controllers for the separated system including stability analyses; numerical simulations are presented in \S \ref{sec:results}, and we conclude the paper in \S \ref{sec:conclude}. 
	\section{Notations and Preliminaries}
\label{sec:prelims}

%
Matrices and vectors are respectively upper- and lower-case  bold-faced letters. The strain field and strain twist vectors are  $\strain \in \reline^6$ and $\twist \in \bb{R}^3$, respectively. Sets, screw stiffness, wrench tensors, and the gravitational vector are upper-case Calligraphic bold-faced characters. Distributed wrench tensors are signified by an overbar, \eg $\bar{\mc{\bm{F}}}$. For a curve  $\abscissa: [0, L]$, where $L$ is the curve's length at time $t$, the robot's configuration is denoted as  $\mathcal{\bm{X}}_t(\abscissa)$. The matrix $\bm{A}$'s Frobenius norm is denoted $\|\bm{A}\|$ while its Euclidean norm is $\|\bm{A}\|_2$. The Lie algebra of the Lie group $\liegroup$ is  $\liealgebra$. The special orthogonal group consisting of corkscrew rotations is $\orthoggroup$. 
The structure's configuration $\conf(\abscissa)$ is  a member of the Lie group $\liegroup$, whose adjoint and coadjoint are respectively denoted $\Adj_{\conf}$, $\Adj_{\conf}^\star$. We remark that these are parameterized by the curve,  $\abscissa$. 
In generalized coordinate, the joint vector of a soft structure is denoted $\gencoord = [\strain_1^\top, \ldots, \strain_{n_\strain}^\top]^\top \in \reline^{6{n_\strain}}$ and $\dot{\gencoord} = [\twist_1^\top, \ldots, \twist_{n_\strain}^\top]^\top \in \reline^{6{n_\strain}}$.  For a roll, pitch and yaw angles $\theta, \phi, \psi$, a typical strain $\strain_i$ or strain twist vector $\twist_i$ takes the forms $[\theta, \phi, \psi, x, y, z]^\top$ and $[\dot{\theta}, \dot{\phi}, \dot{\psi}, \dot{x}, \dot{y}, \dot{z}]^\top$ in our notation.

\subsection{SoRo Configuration}
Our analysis is amenable to many soft robots with one predominantly longer dimension than the other two (see \autoref{fig:pcs_kine}) so that ``thin" Cosserat rod theory~\cite{RubinCosserat} applies. 
Shown in \autoref{fig:pcs_kine}, the inertial frame is the basis triad $(\bm{e}_1, \bm{e}_2, \bm{e}_3)$ and $\conf_r$ is the inertial to base frame transformation. 
%
%
For a cable-driven arm, actuation occurs through the central axis of the robot and at the point $\bar{\abscissa}$ per section. The configuration matrix that parameterizes curve $L_n \in X$ is  $\conf_{L_n}$. The robot's $z$-axis is offset in orientation from the inertial frame by $-90^\circ$ so that a transformation from the base to inertial frames is
\begin{align}
	\conf_r = \left(\begin{array}{cccc}
		0 & -1 & 0 & 0 \\
		1 & 0 & 0 & 0 \\
		0 & 0 & 1 & 0 \\
		0 & 0 & 0 & 1
	\end{array}\right).
	\label{eq:conf_ref}
\end{align} 
\subsection{Continuous Strain Vector and Twist Velocity Fields}
Suppose that $\pos(\abscissa) \in \reline^6$  describes a microsolid's position on the soft body at  $t$  and let $\rot(\abscissa)$ be the corresponding orientation matrix. Let the pose be  $[\pos(\abscissa), \rot(\abscissa)]$. Then, the robot's C-space, parameterized by a curve $g(\cdot): \abscissa \rightarrow \liegroup$, is $
g(\abscissa) = \left(\begin{array}{cc}
	\rot(\abscissa) & \pos(\abscissa) \\
	\bm{0}^\top & 1
\end{array}\right)$.
%
%
Suppose that $\varepsilon(\abscissa) \in \reline^3$ and $\gamma(\abscissa) \in \reline^3$ respectively denote the linear and angular strain components of the soft arm. Then, the arm's strain field is a state vector, $\breve{\xi}(\abscissa) \in \liealgebra$, along the curve $\conf(\abscissa)$ \ie 
$	\breve{\xi}(\abscissa) = \conf^{-1} \partial g/\partial \abscissa\triangleq \conf^{-1} \partial_x \conf.$
%
In the microsolid frame, the matrix and vector representation of the strain state are respectively
%
$	\breve{\xi}(\abscissa) = \left(\begin{array}{cc}
		\hat{\gamma} & {\varepsilon} \\
		\bm{0} & 0
	\end{array}\right) \in \liealgebra, \quad \xi(\abscissa) = \left(\begin{array}{cc}
		\gamma^\top & \varepsilon^\top
	\end{array}\right)^\top \in \reline^6.$
%
Read $\hat{\gamma}$: the anti-symmetric matrix representation of $\gamma$. Read $\breve{\xi}$: the isomorphism mapping the twist vector, $\xi \in \reline^6$, to its matrix  representation in  $\liealgebra$. Furthermore, let $\nu(\abscissa), \omega(\abscissa)$ respectively denote the linear and angular velocities of the curve $g(\abscissa)$. Then, the velocity of $\conf(\abscissa)$ is the twist vector field 
%
$	\breve{\eta}(\abscissa) = \conf^{-1} \partial \conf / \partial t \triangleq \conf^{-1} \partial_t{\conf}.$
%
In the microsolid frame, 
%
$	\breve{\eta}(\abscissa) = \left(\begin{array}{cc}
		\hat{\omega} & {\nu} \\
		\bm{0} & 0
	\end{array}\right) \in \liealgebra, \quad \eta(\abscissa) = \left(\begin{array}{cc}
		\omega^\top & \nu^\top
	\end{array}\right)^\top \in \reline^6.
$

\subsection{Discrete Cosserat-Constitutive PDEs}
The PCS model assumes that  $(\strain_i, \twist_i)$ $i=1,\ldots,N$ robot sections are constant. Spatially spliced along sectional boundaries, the overall strain position and velocity of the entire soft robot is a piecewise sum of the sectional strain field parameters.
	Using d'Alembert's principle, the generalized dynamics for PCS model~\autoref{fig:pcs_kine} under external and actuation loads admits the form~\cite{RendaTRO18}
	\begin{align}
		&\underbrace{\left[\int_0^{L_N} \bm{J}^\top \masscom_a
			\jacob d\abscissa\right]}_{\massmat(\gencoord)}\ddot{\gencoord} + \underbrace{\left[\int_0^{L_N} J^\top \adj_{\jacob\dot{\gencoord}}^\star\masscom_a
			\jacob d\abscissa\right]}_{\bm{C}_1(\gencoord, \dot{\gencoord})}\dot{\gencoord} + \nonumber \\
		&\underbrace{\left[\int_0^{L_N} \jacob^\top \masscom_a
			\dot{\jacob} d\abscissa\right]}_{\bm{C}_2(\gencoord, \dot{\gencoord})}\dot{\gencoord}
		+ \underbrace{\left[\int_0^{L_N} \jacob^\top \bm{\mc{D}}
			\jacob \|\jacob \dot{\gencoord}\|_pd\abscissa\right]}_{\dragmat(\gencoord, \dot{\gencoord})}\dot{\gencoord} \nonumber\\
		& -\underbrace{(1-\rho_f/\rho) \left[\int_0^{L_N} \jacob^\top \bm{\mc{M}} \Adj_{\conf}^{-1} d\abscissa\right]}_{\bm{N}(q)}\Adj_{\conf_r}^{-1}\bm{\mc{G}} - \underbrace{\jacob^\top(\bar{\abscissa}) \bm{\mc{F}}_p}_{\bm{F}(\gencoord)} \nonumber \\
		& - \underbrace{\int_0^{L_N} \jacob^\top \left[\nabla_x\bm{\mc{F}}_i - \nabla_x \bm{\mc{F}}_a + \adj_{\twist_n}^\star \left(\bm{\mc{F}}_i - \bm{\mc{F}}_a\right)\right]}_{\torque(\gencoord)}d\abscissa = 0,
		\label{eq:loads_full_form}
	\end{align}
	for a Jacobian $\jacob(\abscissa)$  (see definition in~\cite{RendaTRO18}), wrench of internal forces $\bm{\mc{F}}_i(\abscissa)$, {distributed} wrench of  actuation loads 
	$\bar{\bm{\mc{F}}}_a(\abscissa)$, and \textit{distributed}  wrench of the applied external forces $\bar{\mc{\bm{F}}}_e(\abscissa)$. 
	The torque and (internal) force are respectively $\bm{M}_k, \bm{F}_k$ for sections $k$; and $\masscom(X)$ is the screw mass inertia matrix, given as $\masscom(X) = \text{diag}\left(\identity_x, \identity_y, \identity_z, \bm{A}, \bm{A}, \bm{A}\right)\rho$ for a body density $\rho$, sectional area $A$, bending, torsion, and second inertia operator $I_x, I_y, I_z$ respectively. 
	In \eqref{eq:loads_full_form}, $\masscom_a = \masscom + \masscom_f$ is a lumped sum of the microsolid mass inertia operator, $\masscom$, and that of the added mass fluid, $\masscom_f$; $dX$ is the length of each section of the multi-robot arm; $\mc{\dragmat}(\abscissa)$ is the drag fluid mass matrix; $\jacob(\abscissa)$ is the Jacobian operator; $\|\cdot\|_p$ is the translation norm of the expression contained therein; $\rho_f$ is the density of the fluid in which the material moves; $\rho$ is the body density; $\mc{\bm{G}}$ is the gravitational vector defined as $\mc{\bm{G}} = \left[0, 0, 0, -9.81, 0, 0\right]^T$; and $\mc{\bm{F}}_p$ is the applied wrench at the point of actuation $\bar{\abscissa}$.

	Suppose that $\bm{z} = \dot{\gencoord}$ and the robot's state at a configuration $\conf$ is $\bm{x} = [\gencoord^\top, \bm{z}^\top]^\top$, then equation \eqref{eq:loads_full_form}  can be appropriately written in standard Newton-Euler (N-E) form as
	\begin{align}
		\begin{split}
			\massmat (\gencoord) \dot{\bm{z}} &+ \left[\bm{C}_1 (\gencoord, \bm{z}) + \bm{C}_2(\gencoord, \bm{z}) + \bm{D}(\gencoord, \bm{z})\right]\bm{z} =  \\ 
			& \qquad \tau(\gencoord) + \bm{F}(\gencoord) + \bm{N}(\gencoord)\Adj_{\conf_r}^{-1} \mc{\bm{G}}.
		\end{split}
		\label{eq:newton-euler}
	\end{align}
	%
%
	\section{Singularly Perturbed Dynamics}
\label{sec:fast-slow}
Seeking a robust response to parametric variations, noise sensitivity, and parasitic small time constants  in the dynamics that increase model order, we separate  system \eqref{eq:newton-euler} into a standard two-time-scale singularly perturbed system consisting of fast-changing (here, $\dot{\bm{z}}_2$) and slow-changing (\ie $\dot{\bm{z}}_1$) sub-dynamics. 
Thus, we write 
\begin{subequations}
	\begin{align}
		\dot{\bm{z}}_1 &= \bm{f}(\bm{z}_1, \bm{z}_2, \perturb, \bm{u}_s, t), \,\, \bm{z}_1(t_0) = \bm{z}_1(0), \,\, \bm{z}_1 \in \reline^{6N} 	\label{eq:spt_standard_f}, \\
		\perturb \dot{\bm{z}}_2 &= \bm{g} (\bm{z}_1, \bm{z}_2, \perturb, \bm{u}_f, t), \,\, \bm{z}_2(t_0)=\bm{z}_2(0), \,\, \bm{z}_2 \in \reline^{6N}
			\label{eq:spt_standard_g}
	\end{align}
	\label{eq:spt_standard}
\end{subequations}
where  $\bm{f}$ and $\bm{g}$ are $\mc{\bm{C}}^n (n \gg0)$ differentiable functions of their arguments, $\perturb > 0$ denotes all small parameters to be ignored\footnote{Restriction to a two-time-scale is not binding and one can choose to expand the system into multiple sub-dynamics across multiple time scales.}, $\bm{u}_s$ is the slow sub-dynamics' control law, and $\bm{u}_f$ is the fast sub-dynamics' controller. 

We assume that the fast feedback law is asymptotically stable (formalized in Assumption \ref{ass:unique_root}) such that it does not modify the open-loop equilibrium manifold of the fast dynamics. Thus, setting $\perturb = 0$ to extract the slow subdynamics (here $\bm{u}_f= 0$) the system dynamics becomes 
\begin{subequations}
	\begin{align}
		\dot{\bm{z}}_1 &= \bm{f}(\bm{z}_1, \bm{z}_2, 0, \bm{u}_s, t), \,\, \bm{z}_1(t_0) = \bm{z}_1(0), \\
		0 &= \bm{g}(\bm{z}_1, \bm{z}_2, 0, 0, t).
	\end{align}
	\label{eq:spt_algebraic}
\end{subequations}
%

\begin{assumption}[Real and distinct root] \label{ass:unique_root}
Equation \eqref{eq:spt_algebraic} has the unique and distinct root $\bm{z}_2 = \bm{\phi}(\bm{z}_1, t)$ (for a sufficiently smooth $\bm{\phi}(\cdot)$)
%
so that 
\begin{align}
	0 = \bm{g}(\bm{z}_1, \bm{\phi}(\bm{z}_1, t), 0, 0, t) \triangleq \bar{\bm{g}}(\bm{z}_1, 0, t) , \,\,  \bm{z}_1(t_0) = \bm{z}_1(0).
\end{align}
The slow subsystem therefore becomes
\begin{align}
	\dot{\bm{z}}_1 = \bm{f}(\bm{z}_1, \bm{\phi}(\bm{z}_1, t), 0, \bm{u}_s, t) \triangleq \bm{f}_s(\bm{z}_1, \bm{u}_s, t).
	\label{eq:spt_slow_model}
\end{align}
\end{assumption}
Assumption \ref{ass:unique_root} is a standard assumption in singular perturbation theory~\cite{SPTBook} and it allows us to isolate the equilibrium manifold of the fast dynamics such that the slow subdynamics takes the form of an algebraic expression. For the fast subdynamics, let us introduce the time scale $T = t/\perturb$, and write the deviation of $\bm{z}_2$ from its isolated equilibrium manifold, $\bm{\phi}(\bm{z}_1, t)$ as $\tilde{\bm{z}}_2 = \bm{z}_2  - \bm{\phi}(\bm{z}_1, t)$. Then,  
\eqref{eq:spt_standard}  becomes
\begin{subequations}
	\begin{align}
		\dfrac{d \bm{z}_1}{dT} &= \perturb \bm{f}(\bm{z}_1, \tilde{\bm{z}}_2 + \bm{\phi}(\bm{z}_1, t), \perturb, \bm{u}_s, t), \\
		\dfrac{d \tilde{\bm{z}}_2}{dT} &= \perturb \dfrac{d \bm{z}_2}{dt} - \epsilon \dfrac{\partial \bm{\phi}}{\partial \bm{z}_1}\dot{\bm{z}}_1,  \\
		%
		&= \bm{g}(\bm{z}_1, \tilde{\bm{z}}_2 + \bm{\phi}(\bm{z}_1, t), \perturb, \bm{u}_f, t) - \perturb \dfrac{\partial \bm{\phi}(\bm{z}_1, t)}{\partial \bm{z}_1}\dot{\bm{z}}_1. 
	\end{align}
\end{subequations}
Setting $\perturb = 0$,  we obtain the algebraic equation 
\begin{align}
	\dfrac{d \bm{\tilde{z}}_2}{dT} &= \bm{g}(\bm{z}_1,  \tilde{\bm{z}}_2 + \bm{\phi}({z}_1, t), 0, \bm{u}_f, t)
\end{align}
with $\bm{z}_1$ frozen to its initial values.

\subsection{Soft Robots' Dynamics Separation}
\label{sec:spt_intro}
The robot's motion can be decomposed into those along the discretized sections' barycenter and those relative to the barycenter based on the discretized Cosserat constant strain assumption. 
Denote the composite mass distribution as a result of microsolid $i'$s barycenter motion as $\masscore_i$. Motions  relative  to $\masscore_i$ can be considered a perturbation,  $\masspert$, so that  $\masspert = \masscom \setminus \masscore$. 
%
%
Examining \eqref{eq:newton-euler}, suppose that the perturbation and core microsolids' indices  are $(L^p_{\min}, L^p_{\max})$ and $(L^c_{\min}, L^c_{\max})$, respectively, where $0 \le L^p_{\min} < L^c_{\min}$,  $L^c_{\max} <  L^p_{\max} \le L$, and $(L^c_{\max} > L^c_{\min}), (L^p_{\max} > L^p_{\min})$.  Then, we can write
\begin{subequations}
	\begin{align}
		&\massmat(\gencoord) = (\massmat^{c}+ \massmat^{p})(\gencoord), \, \Nmat = (\Nmat^{c}+ \Nmat^{p})(\gencoord) ,  \\
		&\forcemat(\gencoord) = (\forcemat^{c} + \forcemat^{p})(\gencoord),   \quad 
		\dragmat(\gencoord) = (\dragmat^{c} + \dragmat^{p})(\gencoord) \\
		&\cormat_1(\gencoord, \dot{\gencoord}) = (\cormat_1^{c} + \cormat_1^{p})(\gencoord, \dot{\gencoord}),  \\
		&\cormat_2(\gencoord, \dot{\gencoord}) = (\cormat_2^{c} + \cormat_2^{p})(\gencoord, \dot{\gencoord})
	\end{align}
	\label{eq:core_pert_sep}
\end{subequations}
where $	\massmat^p = \int_{L^p_{\min}}^{L^p_{\max}} \bm{J}^\top \masscom^{pert} \bm{J} dX$, \text{ and }
$\bm{M}^c = \int_{L^c_{\min}}^{L^c_{\max}} \bm{J}^\top \mathcal{\bm{M}}^{core} \bm{J} dX$,
%
and every other matrix in \eqref{eq:core_pert_sep} is similarly defined. 
%

%
%
Suppose that the respective matrices are in diagonal block form,  the mass inertia matrix in \eqref{eq:core_pert_sep} can be decomposed as (dropping the joint space arguments for ease of readability) 
\begin{align}
	\massmat &= \underbrace{\begin{bmatrix}
		\hcal_{\core} & \zero \\ \zero & \zero
	\end{bmatrix}}_{\massmat^c(\gencoord)} + 
	\underbrace{\begin{bmatrix}
		\zero & \hcal_{\pert}^{\core} \\ {\hcal_{\pert}^{\core}}^\top & \hcal_{\pert}
	\end{bmatrix}}_{\massmat^p(\gencoord)}, 
\end{align}
where each block $\massmat^c(\gencoord)$ and $\massmat^p(\gencoord)$ are invertible (see \cite{LekanSoRoPD}), and by extension $\hcal_\core$ is invertible; $\hcal_\pert^\core$ denotes the decomposed mass of the perturbed sections of the robot relative to the core sections. 

Introducing the change of variables $[\gencoord^\top, \dot{\gencoord}^\top]^\top = [\gencoord^\top, \bm{z}^\top]^\top$, so that the robot's state, $\bm{x} = [\gencoord^\top, \bm{z}^\top]^\top$ decomposes as 
${\gencoord} = [\gencoord_{\core}^\top, \gencoord_{\pert}^\top]^\top, \,\bm{z} = [\bm{z}_{\core}^\top, \bm{z}_{\pert}^\top]^\top,$
where $\bm{x}_\core$ denotes the components of $\bm{x}$ belonging to the fast subsystem and $\bm{x}_\pert$ denotes the components of $\bm{x}$ belonging to the slow subsystem. 
%
Furthermore, let $\bar{\massmat}^p = {\massmat}^p/\perturb$, and let  $\torque = [\torque_{\core}^\top, \torque_{\pert}^\top]^\top$ be the applied torque (control law to be designed). Rewriting \eqref{eq:newton-euler} with the singular perturbation parameter $\perturb = \|\massmat^p \|/\|\massmat^c \|$, we have
\begin{align}
	(\massmat^c + \perturb \bar{\massmat}^p) \dot{\bm{z}} = \bm{s} + \torque,
	\label{eq:system_perturb}
\end{align}
where 
\begin{align}
\bm{s}=	\begin{bmatrix}
		\bm{s}_{\core} \\ \bm{s}_{\pert}
	\end{bmatrix} = \begin{bmatrix}
	\forcemat^c + \Nmat^c \Adj_{\conf_r}^{-1} \bm{\mc{G}} - [\cormat_1^c + \cormat_2^c + \dragmat^c] \bm{z}_\core \\
	\forcemat^p + \Nmat^p \Adj_{\conf_r}^{-1} \bm{\mc{G}} - [\cormat_1^p + \cormat_2^p + \dragmat^p] \bm{z}_\pert 
	\end{bmatrix}. 
\end{align}
Since $\hcal_\core$ is invertible, let  
\begin{align}
	\bar{\massmat}^p = \begin{bmatrix}
		\bar{\massmat}^p_{11} & \bar{\massmat}^p_{12} \\ \bar{\massmat}_{21}^p   & \bar{\massmat}^p_{22}
	\end{bmatrix} \text{ and } \bm{\Delta} = \begin{bmatrix}
		\zero & \zero \\ \bar{\massmat}_{21}^p \hcal_{\core}^{-1}  & \zero 
	\end{bmatrix}, 
\end{align}
then premultiplying both sides of \eqref{eq:system_perturb} by $\identity - \perturb \bm{\Delta}$, and ignoring the squared term in $\perturb$, it can be verified that
\begin{align}
	\begin{bmatrix}
		\hcal_{\core} & \perturb \hcal_{\pert}^{\core} \\
		\zero        & \perturb \hcal_{\pert}
	\end{bmatrix} \begin{bmatrix}
		\dot{\bm{z}}_{\core} \\ \dot{\bm{z}}_{\pert}
	\end{bmatrix} &= \begin{bmatrix}
	{\bm{s}}_{\core} \\ {\bm{s}}_{\pert} - \perturb \bar{\massmat}_{21}^p \hcal_{\core}^{-1} \bm{s}_{\core}
	\end{bmatrix} + \nonumber \\
	&  \begin{bmatrix}
	\torque_{\core} \\ \torque_{\pert} - \perturb \bar{\massmat}_{21}^p \hcal_{\core}^{-1} \torque_{\core}
	\end{bmatrix}.
\end{align}
Rearranging, we must have
\begin{align}
	\begin{bmatrix}
	\hcal_{\core} & \bar{\massmat}^p_{12}\\
	\zero        & \bar{\massmat}_{22}^p
\end{bmatrix} \begin{bmatrix}
	\dot{\bm{z}}_{\core} \\ \perturb \dot{\bm{z}}_{\pert}
\end{bmatrix} &= \begin{bmatrix}
	{\bm{s}}_{\core} \\ {\bm{s}}_{\pert} - \perturb \bar{\massmat}_{21}^p \hcal_{\core}^{-1} \bm{s}_{\core}
\end{bmatrix} + \nonumber \\
&  \begin{bmatrix}
	\torque_{\core} \\ \torque_{\pert} - \perturb \bar{\massmat}_{21}^p \hcal_{\core}^{-1} \torque_{\core}
\end{bmatrix}
\label{eq:soro_perturbed_form}
\end{align}
which is in the standard singularly perturbed form \eqref{eq:spt_standard}.

\subsubsection{Fast subsystem dynamics extraction}

Consider the fast time scale $ T = t/\perturb$, with $dT/dt = 1/\perturb$. The dynamics on this time scale is $\dot{\bm{z}}_{\core} = \dfrac{d \bm{z}_{\core}}{dt} \equiv  \dfrac{1}{\perturb} \dfrac{d \bm{z}_{\core}}{dT} \triangleq \dfrac{1}{\perturb} \bm{z}_{\core}^\prime$ and $\perturb \dot{\bm{z}}_{\pert} =  \bm{z}_{\pert}^\prime$. Rewriting \eqref{eq:soro_perturbed_form}, we find
\begin{align}
	\begin{bmatrix}
		\hcal_{\core} & \perturb \bar{\massmat}^p_{12}\\
		\zero        & \bar{\massmat}_{22}^p
	\end{bmatrix} \begin{bmatrix}
		\bm{z}^\prime_{\core} \\ \bm{z}^\prime_{\pert}
	\end{bmatrix} &= \begin{bmatrix}
		\perturb {\bm{s}}_{\core} \\ {\bm{s}}_{\pert} - \perturb \bar{\massmat}_{21}^p \hcal_{\core}^{-1} \bm{s}_{\core}
	\end{bmatrix} + \nonumber \\
	& \quad \begin{bmatrix}
		\perturb  \torque_{\core} \\ \torque_{\pert} - \perturb \bar{\massmat}_{21}^p \hcal_{\core}^{-1} \torque_{\core}
	\end{bmatrix},
	\label{eq:soro_perturbed_fast}
\end{align}
or, 
\begin{subequations}
	\begin{align}
		&\bm{z}^\prime_{\core} = \perturb\hcal_{\core}^{-1} ({\bm{s}}_{\core} +  \torque_{\core})  -   \hcal_{\core}^{-1}\hcal^{\core}_{\pert} \bm{z}^\prime_{\pert}\\
		& \bm{z}^\prime_{\pert} = \hcal_{\pert}^{-1} ({\bm{s}}_{\pert} - \torque_{\pert}) - \hcal_{\core}^{-1} (\bm{s}_{\core}- \torque_{\core}) 
	\end{align}
	\label{eq:fast_subsys}
\end{subequations}
where the slow variables are frozen on this fast time scale. 

\subsubsection{Slow sub-dynamics}
To extract the slow subdynamics, we let $\perturb \rightarrow 0$ in \eqref{eq:soro_perturbed_form}, so that what is left, i.e,
%
%
\begin{align}
	 \dot{\bm{z}}_{\pert} = \hcal_{\pert}^{-1}({\bm{s}}_{\pert}  + \torque_{\pert})
	\label{eq:slow_subsys}
\end{align}
constitutes the system's slow dynamics, where the fast components are frozen on this slow time scale.
	\section{Hierarchical Controller Synthesis}
\label{sec:hcs}

We seek a \textit{multi-rate feedback backstepping controller} which steer an arbitrary strain configuration $[\gencoord(t)^\top, \dot{\gencoord}(t)^\top]^\top$ at time $t$, to a target point $[\gencoord^{d \top}, \dot{\gencoord}^{d \top}]^\top$. 
We now design nonlinear backstepping controllers for the  separate subsystems in \S \ref{sec:spt_intro}.

\subsubsection{Stability analysis of the fast strain subdynamics}
Let us first consider the velocity component of the fast subdynamics in \eqref{eq:fast_subsys}; this exists on the time scale $t_f \triangleq T \triangleq$. 
Consider the transformation $[\bm{\theta}^\top, \bm{\phi}^\top]^\top = [\gencoord^\top_\core, \bm{z}_{\core}^\top]^\top$ where $\bm{\theta}^\prime = \perturb \zee_\core$. Suppose that we choose the virtual input $\vinput$ such that $\titi^\prime = \vinput$ and let $\gencoord_{\core}^d = [\bm{\xi}_1^d, \ldots, \bm{\xi}_{n_{\xi}}^d]_\core^\top$ be the desired joint space configuration

\begin{theorem}
	The control law 
	$$\gencoord_\core^d(t_f) - \gencoord_\core(t_f) + \gencoord^{\prime d}_\core(t_f)$$
	is sufficient to guarantee an exponential stability of the origin of $\titi^\prime = \vinput$ such that for all $t_f \ge 0$, $\gencoord_\core(t_f) \in S$ for  a compact set $S \subset \reline^{6N}$. That is, $\gencoord_\core(t_f)$ remains bounded as $t_f \rightarrow \infty$.
	\label{thm:fast_control_law_sub1}
\end{theorem}
\begin{proof}
Define the tracking error and corresponding error dynamics as
\begin{subequations}
	\begin{align}
			\error_1 &= \titi  - \gencoord^d_\core, \, \implies	\error_1^\prime = \titi^\prime  - \gencoord^{\prime^d}_\core \triangleq \vinput - \gencoord_\core^{\prime^d}.
	\end{align}
	\label{eq:error_slow_sys}
\end{subequations}
Consider the following candidate Lyapunov function, 
\begin{align}
	\bm{V}_1(\error_1) = \frac{1}{2} \error_1^\top \bm{K}_p \error_1 
\end{align}
where  $\bm{K}_p$ is a diagonal matrix of positive damping (gains). 
Ignoring the templated arguments for ease of readability, for a constant $\gencoord^d_\core$, we must have
	\begin{align}
		\bm{V}_1^\prime&= \error_1^\top \bm{K}_p {\error}^\prime_1 = \error_1^\top \bm{K}_p (\vinput - {\gencoord}^{\prime d}_\core).
	\end{align} 
Set $\vinput =  {\gencoord}^{\prime d}_\core - \error_1$, then
\begin{align}
 \bm{V}_1^\prime = -\error_1 \bm{K}_p \error_1 \le 2 \bm{V}_1.
\end{align}
That is for, $\lim_{t\rightarrow \infty} \error_1(t) = 0$ the control law ${\gencoord}^{\prime d}_\core - \error_1 \triangleq \gencoord_\core^d - \gencoord_\core + \gencoord^{\prime d} $ implies an exponentially stable origin of the subsystem hence satisfying Assumption \ref{ass:unique_root}. 
\end{proof} 

\subsubsection{Stability analysis of the  strain twist subdynamics}
\begin{theorem}\label{thm:fast_subdynamics}
	Under the tracking error $\error_2 = \bm{\phi} - \vinput$ and matrices $(\bm{K}_p, \bm{K}_q) = (\bm{K}_p^\top, \bm{K}_q^\top) >0$, the control input 
	\begin{align}
		\torque_{\core} &= \frac{1}{\perturb} \hcal_{\core} [\gencoord_\core^{\prime \prime d} + \error_1 - 2\error_2 - \bm{K}_q^\top (\bm{K}_q \bm{K}_q^\top)^{-1}\bm{K}_p \error_1]  \nonumber \\
		& \qquad +  \frac{1}{\perturb} \hcal^{\core}_\pert \zee_\pert^\prime -  \bm{s}_{\core} 
		\label{eq:core_controller}
	\end{align}
exponentially stabilizes the fast subdynamics \eqref{eq:fast_subsys}.
\end{theorem}

\begin{proof}
First recall that 
\begin{subequations}
\begin{align}
	\error_1^\prime &= \titi^\prime - \gencoord_{\core}^{\prime d} \triangleq \zee_\core - \gencoord_\core^{\prime d} + (\vinput - \vinput ) \\
	&= (\bm{\phi} - \vinput) + (\vinput - \gencoord_{\core}^{\prime d}) \triangleq \error_2 - \error_1.
\end{align}
\end{subequations}
Now, consider the whole nonlinear fast subsystem \eqref{eq:fast_subsys}.  It follows that 
\begin{align}
	\error_2^\prime &= \bm{\phi}^\prime - \vinput^\prime = \zee_\core^\prime + \error_1^\prime - \gencoord_{\core}^{\prime \prime d} \\
	&= \hcal_{\core}^{-1} \left[\perturb \torque_{\core} + \perturb \bm{s}_\core - \hcal_{\pert}^\core \zee_{\pert}^{\prime}  \right] + (\error_2 - \error_1) - \gencoord_{\core}^{\prime \prime d}. \nonumber
\end{align}

Suppose that we choose the Lyapunov candidate function 
\begin{align}
	\bm{V}_2(\error_1, \error_2) = \bm{V}_1 + \frac{1}{2} \error_2^\top \bm{K}_q \error_2 = \frac{1}{2} [\error_1 \,\, \error_2] \begin{bmatrix}
		\bm{K}_p & \zero \\  \zero & \bm{K}_q
	\end{bmatrix} \begin{bmatrix}
	\error_1 \\ \error_2
\end{bmatrix}, \nonumber
\end{align}
 it can be verified that 
\begin{subequations}
	\begin{align}
		\bm{V}_2^\prime&(\error_1, \error_2)= \error_1^\top \bm{K}_p \error_1^\prime + \error_2^\top \bm{K}_q \error_2^\prime  \\
		&=  \error_1^\top \bm{K}_p (\error_2 - \error_1) + \error_2^\top \bm{K}_q [\hcal_{\core}^{-1}(\perturb \torque_{\core} + \perturb \bm{s}_\core - \nonumber \\
		& \qquad \qquad \hcal_{\pert}^\core \zee_{\pert}^\prime ) + (\error_2 - \error_1) - \gencoord_{\core}^{\prime \prime d}].
	\end{align}
\end{subequations}
Substituting the value of $\torque_{\core}$ in \eqref{eq:core_controller} into the foregoing (and ignoring the templated arguments for ease of readability), we have
\begin{subequations}
	\begin{align}
		\bm{V}_2^\prime &= \error_1^\top \bm{K}_p (\error_2 - \error_1) \nonumber \\
		& \qquad - \error_2^\top \bm{K}_q \left(\error_2 - \bm{K}_q^\top (\bm{K}_q \bm{K}_q^\top)^{-1}\bm{K}_p \error_1 \right) \\
		&= -\error_1^\top \bm{K}_p \error_1 - \error_2^\top \bm{K}_q \error_2 \triangleq -2 \bm{V}_2 \le 0.
	\end{align}
\end{subequations}
Since $\bm{V}_2^\prime$ is negative definite, the equilibrium point $\error_{12} = [\error_1^\top, \error_2^\top]^\top = \zero$ is exponentially stable.
And the controller that satisfies the equilibrium points $[\error_1^\top, \error_2^\top]^\top = \zero$ is  given by \eqref{eq:core_controller} or in simplified form 
\begin{align}
	\torque_{\core} &= \frac{1}{\perturb} \hcal_{\core} [\gencoord_\core^{\prime \prime d} -\tilde{\gencoord}_\core -2 \tilde{\gencoord}_\core^{\prime} - \bm{K}_q^\top (\bm{K}_q \bm{K}_q^\top)^{-1}\bm{K}_p \tilde{\gencoord}_\core] \nonumber \\
	& \qquad \qquad  +  \frac{1}{\perturb} \hcal^{\core}_\pert \zee_\pert^\prime -  \bm{s}_{\core}, \nonumber
\end{align}
where $ \tilde{\gencoord}_\core= \gencoord_\core -\gencoord_\core^d$ and $\tilde{\gencoord}^\prime_\core= \gencoord^\prime_\core -\gencoord^{\prime d}_{\core}$.
	%
	On the fast subsystem, the control input value when the perturbed parameters are frozen is 
	\begin{align}
		\torque_{\pert} = \bm{s}_\pert - \hcal_{\pert} \zee_\pert^\prime - \hcal_\pert \hcal_{\core}^{-1}(\bm{s}_\core - \torque_{\core})
	\end{align}
	where the variables $\bm{s}_\pert$, $\hcal_\pert, \zee_\pert^\prime$ are frozen.
\end{proof}

\subsubsection{Stability analysis of the slow strain subdynamics}
For the slow subsystem \eqref{eq:slow_subsys}, let 
$\error_3 = \zee_{\pert} - \vinput$ so that $\dot{\error}_3 = \dot{\zee}_{\pert} - \dot{\vinput}$. It follows that 
\begin{subequations}
	\begin{align}
		\dot{\error}_3 &= \dot{\zee}_{\pert} - \ddot{\gencoord}_\core^d + (\error_2 - \error_1), \\
		&= \hcal_\pert^{-1} (\bm{s}_\pert + \torque_{\pert}) -  \ddot{\gencoord}_\core^d + (\error_2 - \error_1) .
	\end{align}
	\label{eq:slow_subsynamics}
\end{subequations}
%
%
\begin{theorem}
	The control law 
	\begin{align}
		\torque_\pert = \hcal_{\pert}(\error_1 - \error_2 - \error_3 + \ddot{\gencoord}^d_\core) - \bm{s}_\pert
		\label{eq:slow_controller_thm}
	\end{align}
	exponentially stabilizes the slow subdynamics.
\end{theorem}
\begin{proof}
	Consider the Lyapunov function candidate
	\begin{align}
		\bm{V}_3(\error_3) = \frac{1}{2} \error_3^\top \bm{K}_r \error_3  \text{ where } \bm{K}_r = \bm{K}_r^\top > 0.
	\end{align}
	It follows that 
	\begin{subequations}
		\begin{align}
			&\dot{\bm{V}}_3(\error_3) = \error_3^\top \bm{K}_r \dot{\error}_3  \\
			&= 
			\error_3^\top \bm{K}_r \left[\hcal_\pert^{-1} (\bm{s}_\pert + \torque_{\pert}) -  \ddot{\gencoord}_\core^d + \error_2 - \error_1\right]. 
		\end{align}
	\end{subequations}
	Substituting $\torque_\pert$ in \eqref{eq:slow_controller_thm}, it can be verified that
	\begin{align}
		\dot{\bm{V}}_3(\error_3) = \error_3^\top \bm{K}_r \error_3 \triangleq -2 {\bm{V}}_3(\error_3) \le 0.
		\label{eq:slow_controller}
	\end{align}
	Hence, the controller \eqref{eq:slow_controller_thm} stabilizes the slow subsystem.
\end{proof}

\subsubsection{Stability of the singularly perturbed interconnected system}
Let $\lyapweight = (0, 1)$ and consider the composite Lyapunov function candidate $\Sigma(\zee_\core, \zee_\pert)$ as a weighted combination of $\bm{V}_2$ and $\bm{V}_3$ \ie,  
\begin{align}
	\bm{\Sigma}(\zee_\core, \zee_\pert) = (1-\lyapweight) 	\bm{V}_2(\zee_\core) + \lyapweight \bm{V}_3(\zee_\pert), \, 0 < \lyapweight <1.
\end{align}
It follows that,
%
	\begin{align}
		\dot{\bm{\Sigma}}&(\zee_\core, \zee_\pert) = (1-\lyapweight)[ \error_1^\top \bm{K}_p \dot{\error}_1 + {\error}_2^\top \bm{K}_q \dot{\error}_2]  + \lyapweight {\error}_3^\top \bm{K}_r \dot{\error}_3, \nonumber \\
		& = -2(\bm{V}_2 +  \bm{V}_3) + 2 \lyapweight \bm{V}_2 \le 0 
	\end{align}
%
which is clearly negative definite for any $\lyapweight \in (0, 1)$. Therefore, 
we conclude that the origin of the singularly perturbed system is asymptotically stable under the control laws.
\begin{align}
	\torque(\zee_\core, \zee_\pert) = (1-\lyapweight) \torque_{\core} + \lyapweight \torque_{\pert}.
\end{align}
	\section{Numerical Results}
\label{sec:results}

\subsection{System Setup}
\label{sec:setup}
We replicate the parameters of~\cite{LekanSoRoPD} with tweaks to accommodate our layered control method. As seen in \autoref{fig:pcs_kine}, the tip load acts on the $+y$-axis in the robot's base frame so that the tip wrench applied at $\bar{X}=L$, can be expressed as
\begin{align}
	\bm{\mc{F}}_p = \text{diag}\left(\bm{R}^\top(L), \bm{R}^\top(L)\right)
	\left(\begin{array}{cccc}
			\bm{0}_{3\times 1} & 0 & 10 & 0
		\end{array}\right)^\top 
	\label{eq:tip_forces}
\end{align}
where $\bm{R}(L)$ is the first $3\times  3$ block submatrix of \eqref{eq:conf_ref}. 
We use $\bm{\mc{F}}_p^y$ to represent the tip load acting along the $+y$ direction in what follows. 
Given the geometry of the robot, we chose a drag coefficient of $0.82$ (a Reynolds number of order $104$) for underwater operations. We set the Young's modulus as $\bm{E} = 110 kPa$ and the shear viscosity modulus to $3 kPa$. The bending second inertia momenta are $I_y = I_z =\pi r^4/4$ while the torsion second moment of inertia is  $I_x=\pi r^4/2$ for $r=0.1 m$, the arm's radius -- uniform across sections. The arm length is $L=2m$. We assume a (near-incompressible) rubber material makes up the robot's body with Poisson ratio $0.45$; the mass is chosen as $\bm{\mc{M}} = \rho \, \cdot \,  \text{diag}([I_x, I_y, I_z, A, A, A])$ for a cylindrical soft shell's nominal density of  $\rho=2,000 kg m^{-3}$ as used in~\cite{RendaTRO18}; the cross-sectional area $A=\pi r^2$ so that $I_x=\pi r^4/2$. The drag screw stiffness matrix $\dragmat$ in \eqref{eq:newton-euler} is a function of each section's geometry and hydrodynamics so that $\dragmat = -\rho_w \nu^T \nu \bm{\breve{D}} \nu/|\nu|$ where $\rho_w$ is the  water density set to $997 kg/m^3$, and $\bm{\breve{D}}$ is the tensor that models the geometry and hydrodynamics factors in the viscosity model (see \cite[\S II.B, eq. 6]{RendaTRO18}). The curvilinear abscissa, $X \in [0, L]$ was discretized into $41$ microsolids per section. For integrating the system duynamics, we adopt a Runge-Kutta-Fehlberg (RKF) integrator implemented in PyTorch.  For every discretized Cosserat piece in our evaluations, we further divided each piece into 13 segments to accommodate the PCS algorithm's modeling precision. Unlike the extremely fine resolution of segments discretization ($\approx 64$) in~\cite{LekanSoRoPD}, we found that this coarse segmentation scheme does not diminish simulation fidelity in all our testing.

\subsection{Deployment and Discussion}
We asynchronously deployed the slow and fast controllers on both subsystems using two separate threads: the slow controller \eqref{eq:slow_controller} was deployed on a host CPU while the fast controller \eqref{eq:core_controller} was deployed in parallel PyTorch~\cite{pytorch} thread on a CUDA-capable GPU. The slow subsystem state, $\zee_\pert$, and control $\torque_\pert$ are retrieved from a Linux named pipe within the faster subsystem's thread. Computation on the slow subsystem are frozen when computing $\zee_\core$ and  $\torque_{\core}$ in the fast subsystem thread.  We now report two numerical experiments (for the sake of conciseness) to validate our new scheme. Further testing and evaluation are available in the online code repository.

\begin{figure}[tb!]
	\centering 
	\includegraphics[width=\columnwidth]{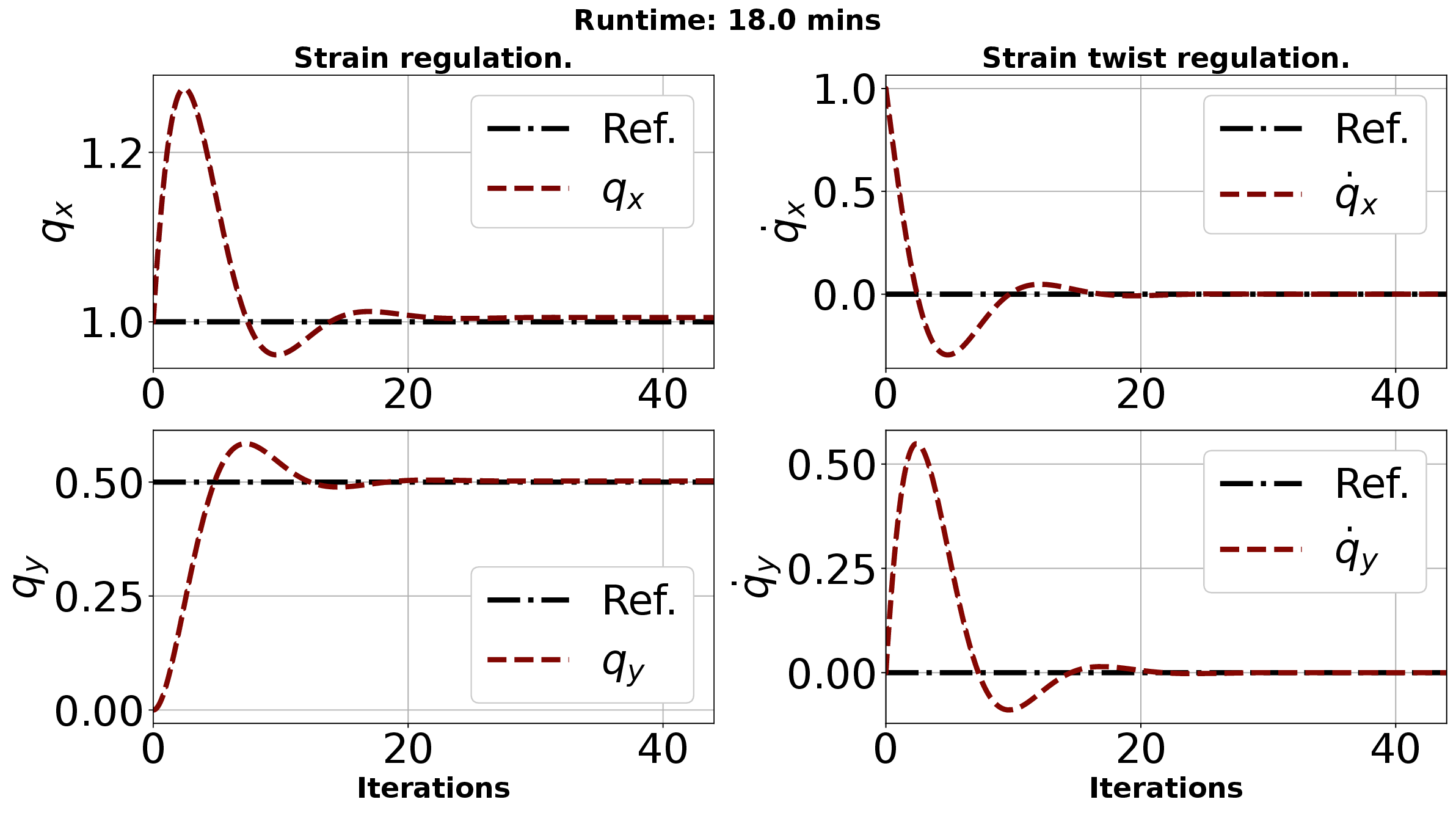}
	\caption{Backstepping control on the singularly perturbed soft robot system with 10 discretized pieces, divided into 6 fast and 4 slow pieces. For a tip load of $\bm{\mc{F}}_p^y=10\,N$, the backstepping gains were set as $\gain_p = 10$, $\gain_d = 2.0$ for a desired joint configuration $\strain^d = [0, 0, 0, 1, 0.5, 0]^\top$ and ${\twist}^d = \zero_{6\times 1}$ that is uniform throughout the robot sections. }
	\label{fig:two_ax_uncorr}
\end{figure}

In a two-axes strain regulation control experiment, we discretized the continuum robot described in \S \ref{sec:setup} into 6 pieces. The fast and slow subdynamics were divided up as 4 and 2 pieces, respectively. The goal is to have the continuum strain along the $+x$ and $+y$ directions as $1.0$ and $0.5$ respectively whilst every other axis is kept at zero under a tip load $10$ Newtons. We set gains $\gain_p = 5$ and $\gain_d = 0.5$. \autoref{fig:two_ax_uncorr} shows the strain and strain twist stabilization results under a total runtime of 18 minutes. As seen, the system reaches steady state across all axes of interest. We remark that this whole body control scheme takes tens of hours for a typical soft robot (later reported in Table \ref{tbl:comparison}). This experiment confirms our hypothesis that dynamics decomposition and nonlinear control aids fast strain regulation. 

Our second experiment employs a PCS scheme with 10 discretized Cosserat sections --- with six fast and flow sub-dynamics, respectively. Under a tip load $\bm{\mc{F}}_p^y=10\,N$, and backstepping gains $\gain_p = 10$, $\gain_d = 2.0$ we aim for desired strain states $\strain^d = [0, \pi/3, \pi, 0.85, 0.5, \pi/4]^\top$ and twist states ${\twist}^d = \zero_{6\times 1}$. \autoref{fig:five_ax_uncorr} shows we reached equilibrium in less than 20 iterations of running the RKF scheme within $25$ minutes. We found the strain states reached steady state within 25 minutes. The results are shown in \autoref{fig:five_ax_uncorr}.
\begin{figure*}[tb!]
	\centering 
	\includegraphics[width=\textwidth]{figures/SPT/five_axes_control.eps}
	\caption{Backstepping control on the singularly perturbed soft robot system with $10$ pieces   $4$ slow and $6$ fast sections.}
	\label{fig:five_ax_uncorr}
\end{figure*}

We further compared the time to reach steady state in our hierarchical control scheme versus a previous work~\cite{LekanSoRoPD} that employed a PD single-layer control law scheme. Here, we employ a similar amount of discretized Cosserat sections and segments in the hierarchical controller (13 segments per sections); while the PD controller employed 41 segments per section. An equal amount of tip load, \ie $10N$ was employed in all experiments. Computations were carried out on an 80GB A100 CUDA-capable NVIDIA GPU for the single layer PD, and fast  subdynamics' controllers. The slow subsystem was executed on the host CPU thread. Table \ref{tbl:comparison} elucidates our results. In all experiments, we found that the hierarchical scheme was significantly faster in reaching equilibrium whilst preserving whole-body strain regulation compared against the PD strain regulation law. 
 \begin{savenotes}
 	\begin{table}[tb!]
 		\centering
 		\begin{tabular}{|p{0.5cm}|c|p{0.4cm}|p{1.5cm}|c|p{0.5cm}|c|p{0.3cm}|r|}
 			\hline
 			\multicolumn{3}{|c|}{\footnotesize{Pieces}}  & \multicolumn{2}{|c|}{\footnotesize{ Runtime (mins)}} \\ \hline 
 			Total & Fast & Slow  & \footnotesize{Hierarchical SPT (mins)} & \footnotesize{Single-layer PD control (hours)}  \\ \hline 
 			6  & 4 & 2 &  $18.01$  & $51.46$  \\ \hline
 			8 & 5 & 3 & $30.87$ & $68.29$  \\ \hline
 			10 & 7 & 3 &  $32.39$   & $107.43$  \\ \hline 
 		\end{tabular}
 	\caption{\footnotesize{Time to Reach Steady State.}}
 	\label{tbl:comparison}
 	\end{table}
 \end{savenotes}
 
	\section{Conclusion}
\label{sec:conclude}
In the quest towards the adoption of soft robots in everyday automation processes, we identified that the long processing times for computing models and controllers/policies is a significant drawback.  To circumvent this, we introduced a singularly perturbed technique for decomposing system dynamics to a fast and slower subdynamics, respectively. Stabilizing nonlinear backstepping controllers were introduced to the respective subdynamics to further improve computation times. The fast part was controlled at a finer resolution while the slower part was controlled at a more coarse resolution, with the overall scheme executed in a decentralized fashion.  We found that our results do not merely regulate particulate strain states but also achieve desired equilibrium faster and better compared to other reported schemes. Our approach takes a further step towards replicating embodied intelligence~\cite{RendaNaturePhysics} in soft robots that mimic the behavior of living matter by engrossing  hierarchy layers in soft robots' dynamics and control computational schemes. 
	\bibliographystyle{plainnat}
	\bibliography{SoRoBC}
	%
	%
	\numberwithin{equation}{section}	
	\setcounter{equation}{0}
	\setcounter{lemma}{0}
	\setcounter{theorem}{0}
\end{document}